%% file: onecolumn.tex
\documentclass[10pt,draftcls,onecolumn]{IEEEtran}
\ifCLASSINFOpdf
\else
   \usepackage[dvips]{graphicx}
\fi
\usepackage{url}
\usepackage{graphicx}
\usepackage{amsmath}
\usepackage{comment}
\usepackage[utf8]{inputenc}
\usepackage[english]{babel}
\usepackage{color}
\usepackage{framed}
\usepackage{amsmath,amssymb,amsfonts}

\usepackage{graphicx}
\usepackage{upgreek}
\usepackage{textcomp}
\usepackage{subfigure}
\usepackage{url}

\usepackage{cite}

\input{symbols.tex}
\usepackage{bbold}
\usepackage{ifpdf}
\usepackage{times}
\usepackage{layout}
\usepackage{float}
\usepackage{afterpage}
\usepackage{amsmath}
\usepackage{amstext}
\usepackage{amssymb,bm}
\usepackage{latexsym}
\usepackage{color}
\usepackage{flexisym}
\usepackage{kantlipsum}
\usepackage{graphicx}
\usepackage{amsmath}
\usepackage{amsthm}
\usepackage{graphicx}

\usepackage{booktabs}
\usepackage{multicol}

\usepackage{enumitem}

\usepackage[switch]{lineno}
\usepackage[named]{algo}
\usepackage[noend]{algpseudocode}
\usepackage{algorithm}

\makeatletter
\newcommand*{\rom}[1]{\expandafter\@slowromancap\romannumeral #1@}
\makeatother

\begin{document}
\setlength{\abovedisplayskip}{3pt}
\setlength{\belowdisplayskip}{3pt}

\title{Collaborative Automotive Radar Sensing via Mixed-Precision Distributed  Array Completion 
}

\author{
Arian Eamaz, Farhang Yeganegi$^{\star}$, Yunqiao Hu$^{\star}$, Mojtaba Soltanalian, and Shunqiao Sun   

\vspace{-20pt}
\thanks{
Arian Eamaz, Farhang Yeganegi and Mojtaba Soltanalian are with the ECE Department, University of Illinois at Chicago, Chicago, IL 60607 USA.
}
\thanks{
Yunqiao Hu and Shunqiao Sun are with the Department of Electrical and Computer Engineering, University of Alabama, Tuscaloosa, AL 35487 USA.}
\thanks{This work was supported in part by the U.S. National Science Foundation Grants CCF-1704401, CCF-2153386 and ECCS-2340029. (\emph{Corresponding author: Arian Eamaz})}
\thanks{$^\star$ Equal contribution, 
the names were arranged in alphabetical order.}
}
\markboth{
}
{Shell \MakeLowercase{\textit{et al.}}: Bare Demo of IEEEtran.cls for IEEE Journals}
\maketitle

\begin{abstract}
This paper investigates the effects of coarse quantization with mixed precision on measurements obtained from sparse linear arrays, synthesized by a collaborative automotive radar
sensing strategy. The mixed quantization precision significantly reduces the data amount that needs to be shared from radar nodes to the fusion center for coherent processing. We utilize the low-rank properties inherent in the constructed Hankel matrix of the mixed-precision array, to recover azimuth angles from quantized measurements. Our proposed approach addresses the challenge of mixed-quantized Hankel matrix completion, allowing for accurate estimation of the azimuth angles of interest. To evaluate the recovery performance of the proposed scheme, we establish a quasi-isometric embedding with a high probability for mixed-precision quantization. The effectiveness of our proposed scheme is demonstrated through numerical results, highlighting successful reconstruction.
\end{abstract}


\setlength{\abovedisplayskip}{3pt}
\setlength{\belowdisplayskip}{3pt}

\section{Introduction}
\label{intro}
Millimeter wave (mmWave) automotive radars at $77$ GHz are highly reliable in all weather environments \cite{sun2020mimo}. Benefiting from multiple-input multiple-output (MIMO) radar
technology, mmWave radars can synthesize virtual arrays with large aperture sizes using a small number of transmit and receive antennas \cite{sun2020mimo}. To reduce the hardware cost, sparse arrays synthesized by MIMO radar technology have been widely adopted in automotive radar. Collaborative sensing offers numerous benefits for object detection and radar imaging over single automotive radar sensor \cite{Collaborative_Radar_EuRAD_2023}. For example, enhancing signal gain, increasing reliability, adaptability, and greater spatial diversity\cite{nanzer2021distributed,Batu_distribued_radar_network_2022}. In automotive sensing tasks, such as direction-of-arrival (DOA) estimation, collaborative sensing involves coherently integrating measurements from subarrays of each radar to synthesize an array with a larger aperture, thereby enhancing angle resolution and increasing Signal-to-Noise Ratio (SNR) at the same time\cite{zhang2023direction}. Hence, there is significant interest in developing novel radar sensing methods within the collaborative sensing framework\cite{Collaborative_Radar_EuRAD_2023}. However, sharing the beam vectors collected from collaborative radar sensing with the fusion center, which contains high-resolution data, can pose practical difficulties and lead to the wastage of communication bandwidth. To circumvent this issue, quantizing the beam vectors appears to be a highly efficient solution.

Conventional high-resolution quantization methods typically demand many quantization levels, resulting in heightened power consumption, elevated manufacturing expenses, and diminished analog-to-digital converters (ADCs) sampling rates. In search of alternative solutions, researchers have delved into coarse quantization scenarios, such as one-bit quantization, where signals are compared with a fixed threshold at ADCs, producing binary outputs. This approach facilitates high-rate sampling while mitigating implementation costs and energy consumption compared to multi-bit ADCs. One-bit ADCs hold significant promise in 
MIMO systems \cite{kong2018multipair}.

However, utilizing a pure one-bit ADC system can encounter challenges, such as significant rate loss in high SNR scenarios and dynamic range issues, where a strong target may overshadow a weaker one \cite{mo2014high}. To mitigate these concerns, a mixed-ADC architecture has been proposed \cite{liang2016mixed}, where most receive antenna outputs are sampled by one-bit ADCs, while high-resolution ADCs sample a few. Recent studies \cite{zhang2024novel} have explored the optimal arrangement of ADCs based on the Cramér-Rao Bound (CRB) of acquired measurements, suggesting that evenly distributing high-resolution ADCs along the edges of linear arrays enhances the CRB for Direction of Arrival (DoA) estimation. However, even multi-bit quantization with more than 4 bits in practical scenarios can effectively mimic a high-resolution setting \cite{mollen2017achievable}. In \cite{eamaz2023automotive}, we outlined the construction of a Hankel matrix utilizing one-bit data obtained from Sparse Linear Arrays (SLA). Through both numerical simulations and theoretical analysis, we demonstrated that utilizing a one-bit Hankel matrix can achieve successful target detection performance with a high probability.

In this paper, we introduce a collaborative radar sensing system to synthesize a distributed sparse array with large aperture where most beamvectors are quantized as one-bit data, while the remaining data is quantized as multi-bit. 
We construct a Hankel matrix using the acquired mixed-precision data, which contains missing entries. Consequently, the problem of target azimuth estimation is transformed into completing the Hankel matrix from quantized data.
It has been demonstrated that when the scalar parameter of uniform dithers is designed to dominate the dynamic range of measurements, multi-bit scalar quantization simplifies to a one-bit comparator\cite{10414385}. Taking advantage of the properties of uniform dithering, which are adept at canceling quantization effects in expected values, the $\ell_1$-norm distance between the matrix and its estimate finds an upper bound with a certain probability. 

Our numerical findings showcase the efficacy of incorporating mixed ADCs alongside a well-suited dithering scheme. This combination efficiently utilizes the low-resolution dither samples during the reconstruction phase. As a result, we achieve significant outcomes, including precise detection of target azimuth locations with enhanced resolution.

\underline{\emph{Notation}}: Throughout this paper, we use bold lowercase and bold uppercase letters for vectors and matrices, respectively. We represent a vector $\mathbf{x}$ and a matrix $\mbX$ in terms of their elements as $\mathbf{x}=[x_{i}]$ and $\mathbf{X}=[X_{i,j}]$, respectively. The sets of real and complex numbers are denoted by $\mathbb{R}$ and $\mathbb{C}$, respectively. The vector/matrix transpose is denoted by $(\cdot)^{\top}$. Given a scalar $x$, we define the operator $(x)^{+}$ as $\max\left\{x,0\right\}$.
The nuclear norm of a matrix $\mbX\in \mathbb{C}^{n_1\times n_2}$ is denoted by $\left\|\mbX\right\|_{\star}=\sum^{r}_{i=1}\sigma_{i}$ where $r$ and $\left\{\sigma_{i}\right\}^{r}_{i=1}$ are the rank and singular values of $\mbX$, respectively. The Frobenius norm of a matrix $\mathbf{X}\in\mathbb{C}^{n_1\times n_2}$ is defined as $\|\mathbf{X}\|_{\mathrm{F}}=\sqrt{\sum^{n_1}_{r=1}\sum^{n_2}_{s=1}\left|x_{rs}\right|^{2}}$, where $x_{rs}$ is the $(r,s)$-th entry of $\mathbf{X}$. In the real case, we also define $\|\mbX\|_{\mathrm{max}}=\sup_{i,j}|X_{i,j}|$. The $\ell_1$-norm for a matrix $\mbX$ means $\|\mbX\|_1=\|\operatorname{vec}(\mbX)\|_1$. 
The operator $\operatorname{diag}\left(\mathbf{b}\right)$ denotes a diagonal matrix with $\{b_{i}\}$ as its diagonal elements.  The set $[n]$ is defined as $[n]=\left\{1,\cdots,n\right\}$. The Hadamard (element-wise) product is $\odot$.
The notation $x \sim \mathcal{U}_{[a,b]}$ means a random variable drawn from the uniform distribution over the interval $[a,b]$. 
The memoryless scalar quantizer is denoted by
$\mathcal{Q}_{_{\Delta}}: \mathbb{R} \rightarrow \mathcal{A}_K$,
where $\Delta$ is the resolution parameter and $\mathcal{A}_K$ is the finite alphabet set given by
\begin{equation}
\label{a2}
\mathcal{A}_K:=\{ \pm  \frac{k \Delta}{2}:  k \in [K]\}.
\end{equation}
The relation between the number of bits $B$ and $K$ is given by $B = \log_2(K)+1$. When we introduce a uniform dither generated as $\tau\sim \mathcal{U}_{\left[-\frac{\Delta}{2},\frac{\Delta}{2}\right]}$, to the input signal of the quantizer, the resulting quantization process is termed as \emph{uniform quantization}. This process can be defined as follows:
\begin{equation}
\label{a1}
\mathcal{Q}_{_{\Delta}}\left(x\right)=\Delta\left(\left\lfloor\frac{x+\tau}{\Delta}\right\rfloor+\frac{1}{2}\right).
\end{equation}

\section{Collaborative Automotive Radar Sensing with Mixed-Precision ADCs} 
This section briefly elucidates 
the collaborative radar sensing system. Subsequently, we introduce our scheme of mixed-precision uniform quantization. This method involves sampling certain measurements and quantizing the remainder using a multi-bit scheme, guided by the design of scalar parameters for uniform dithers. 


\subsection{Coherent Automotive Radar Network with Distributed Sparse Array and Mixed Quantization Precision}
\label{sec_1}

\begin{figure*}
\centering
\subfigure[]{\centering{\includegraphics[width=2.5 in]{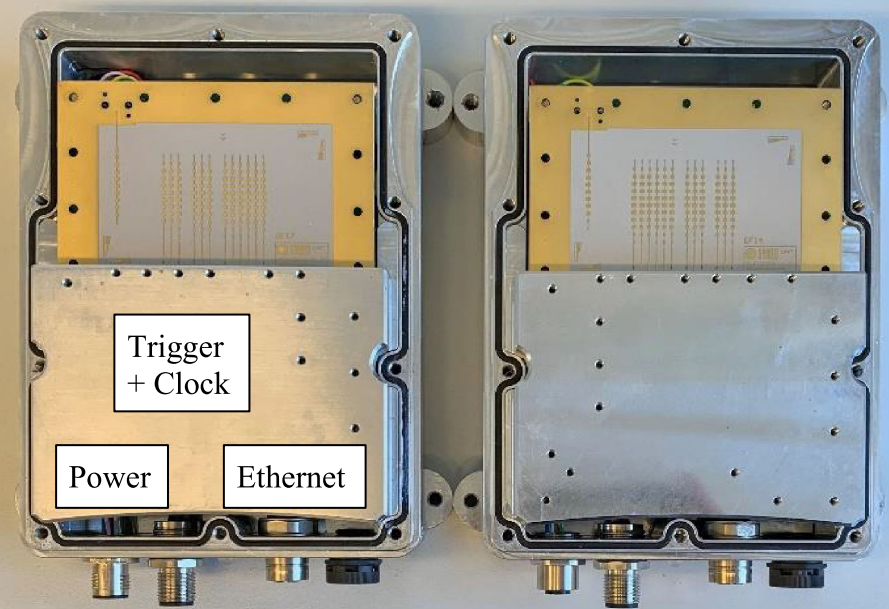}}}\qquad
\subfigure[]{\centering{\includegraphics[width=2.5 in]{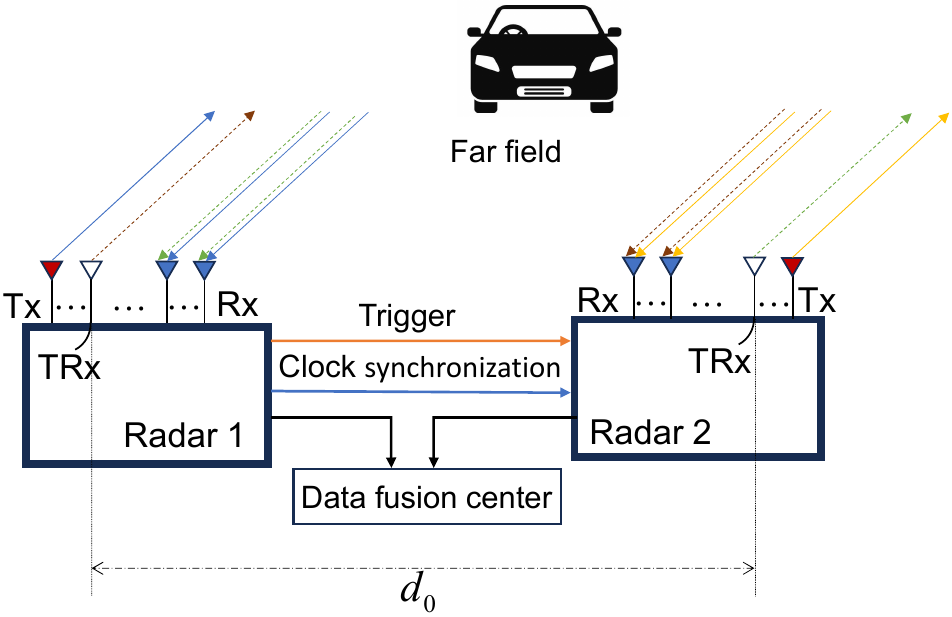}}}\qquad
\subfigure[]{\centering{\includegraphics[width=2.5 in]{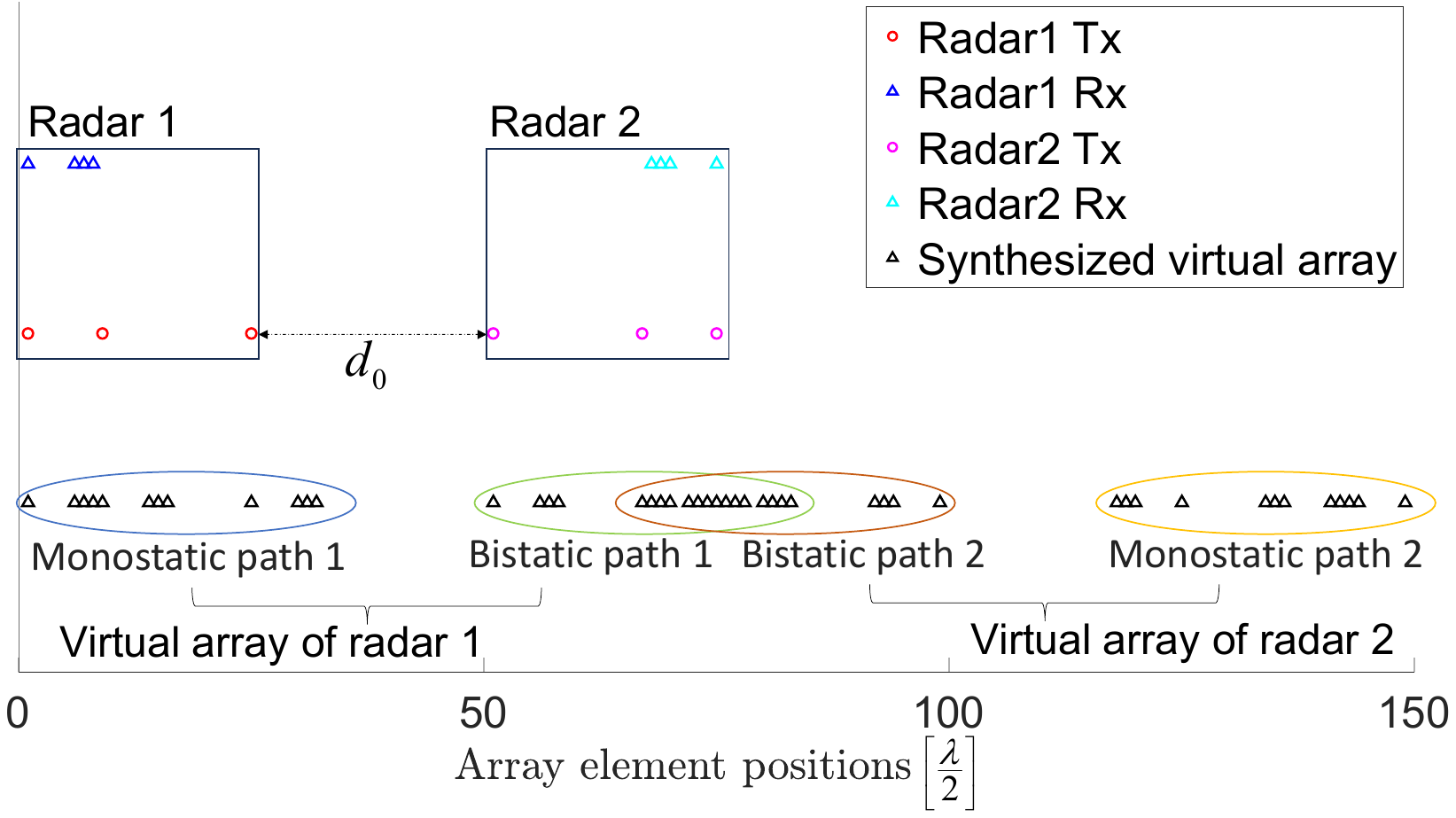}}}
\caption{(a) Collaborative sensing hardware demo consisting of two MIMO FMCW radar units  \cite{gottinger2021coherent}; (b) Simplified illustration of a collaborative sensing system with two radar units, each with 3 TXs and 4 RXs. The two radars are clock synchronized. The beamvectors from each radar are shared to data fusion center for coherent processing. (c) Example of an automotive collaborative radar with $14$ physical array elements and the synthesized virtual distributed sparse array of $48$ elements and aperture of $76\lambda$.
}
\vspace{-10pt}
\label{collab_sensing}
\end{figure*}

Typical automotive radar sensor for Advanced Driver Assistance Systems (ADAS) functions comprises a single radar chipset with 3 transmit and 4 receive antennas. This configuration leads to a restricted array aperture, consequently limiting the sensor's spatial resolution \cite{sun2020mimo}.
We consider an automotive radar network consisting of multiple low-end automotive radars for enhanced array aperture through coherent collaborative sensing. One example of such a radar network with two automotive radars is illustrated in Fig.~\ref{collab_sensing} (a) \cite{gottinger2021coherent}. As shown in Fig.~\ref{collab_sensing} (b), the collaborative sensing consists of a multistatic radar system consisting of two automotive radars. The two radars are clock synchronized. The beamvectors from each radar are shared to data fusion center for coherent processing. 
Each radar receives signal from both monostatic and bistatic paths. The monostatic path denotes the line-of-sight trajectory traversed by the radar signal from the transmitter to the target, and back to the receiver, depicted by the solid blue and yellow lines. The bistatic path signifies the trajectory of the signal from the transmitter of one radar to the target, and then back to the receiver of another one, depicted by the green and brown dotted line in Fig.~\ref{collab_sensing} (b). We suppose the far-field model still holds since the two radars are parallel and the ratio of the separation $d_{0}$ between the two radars and the target distance satisfies the far-field condition. 

As depicted in Fig.~\ref{collab_sensing} (c), each radar features $M_t=3$ Tx and $M_r=4$ Rx antennas, arranged on a grid with a grid size equal to $\lambda/2$, where $\lambda$ is the wavelength. Each automotive radar knows the orthogonal slow-time phase codes of both radars that utilize frequency-modulated continuous-wave (FMCW). Leveraging the orthogonality of waveforms, at each automotive radar receiver, following the MIMO radar concept, the raw two-dimensional (2D) data (fast time $\times$ slow time) with high-precision ADCs corresponding to each Tx waveform can be firstly extracted. The multistatic radar system first performs local range-Doppler processing via 2D fast Fourier transform (FFT). Each radar yields $M_tM_r = 12$ virtual array elements from monostatic path, depicted in Fig.~\ref{collab_sensing} (c) as ``Monostatic path 1'' and ``Monostatic path 2''. Additionally, each radar yields another virtual array with $M_tM_r = 12$ elements from bistatic path, as depicted in Fig.~\ref{collab_sensing} (c) as ``Bistatic path 1'' and ``Bistatic path 2''. The spatial separation $d_{0} = 25\lambda/2$ to accomodate the two radomes.
Each radar now samples these $2M_tM_r = 24$ virtual elements with mixed ADCs and sends them to the fusion center, where a distribued sparse array of $48$ elements with mixed ADCs precision is constructed. Note the data cube of beamvector that each automotive radar sends to the fusion center is of size $N_{\rm fast} \times N_{\rm slow} \times 2M_tM_r$. Thus, mixed ADCs help save the communication bandwidth between each low-end radar and the data fusion center. 


As Fig.~\ref{collab_sensing} (c) shows, with the knowledge of all the orthogonal waveform codes from both radars, each radar could formulate a virtual array with $2 M_t M_r$ elements, and a large distributed sparse array with $4 M_{t} M_{r}$ elements could be synthesized in the data fusion center by  combining the beamvectors sent from these two radars. In general, the element locations of sparse linear array can be considered a subset of a uniform linear array (ULA) antenna positions. Without loss of generality, let the antenna positions of an $M$-element ULA be $\left \{ kd \right \}$, $k=0,1,\cdots,M-1$, where $d=\lambda/2$ is the element spacing with wavelength $\lambda$. As targets are first separated in range-Doppler domain, there is few targets that need to be resolved in angular domain \cite{sun20214d}. Assume there are $P$ uncorrelated far-field target sources in the same range-Doppler bin, and $P$ is sparse. The impinging signals on the ULA antennas are corrupted by additive white Gaussian noise with the variance of $\sigma^2$. For the single-snapshot case, 
the received signal from a ULA as
\begin{align}
\mathbf{x} =  \mathbf{A}\mathbf{s} +  \mathbf{n},
\label{signal_model}
\end{align}
where 
\begin{equation}
\begin{aligned}
\mathbf{x} &= \left[x_{1},x_{2},\dots,x_{M}\right]^{\top},\\  \mathbf{A} &= \left[\mathbf{a}\left(\theta_{1}\right),\mathbf{a}\left(\theta_{2}\right)\dots,\mathbf{a}\left(\theta_{P}\right)\right]^{\top},
\end{aligned}
\end{equation}
with 
$\mathbf{a}\left(\theta_{k}\right) = \left [1,e^{\mathrm{j}2\pi \frac{d\sin\left(\theta_{k}\right)}{\lambda}},\dots , e^{\mathrm{j}2\pi \frac{\left (M-1\right)d\sin \left(\theta_{k}\right)}{\lambda}}\right]^{\top}$,
for $k = 1,\cdots,P$ and $\mathbf{n} = \left[n_{1},n_{2},\dots,n_{M}\right]^{\top}$. We utilize a 1D distributed virtual SLA synthesized by collaborative radar sensing techniques \cite{gottinger2021coherent} with $M_{t}$ transmit antennas and $M_{r}$ receive for each radar. The SLA has $4 M_{t}M_{r} < M$ elements while retaining the same aperture as ULA. Denote the array element indices of ULA as the complete set $\left[M\right]$, the array element indices of SLA can be expressed as a subset ${\Omega^{\prime}}\subset\left[M\right]$ with $\left|\Omega^{\prime}\right|=4M_tM_r$. Thus, the signals received by the SLA can be viewed as partial observations of $\mathbf{x}$, and can be expressed as $\bar{\mbx} = {\bf{m}}_{\Omega^{\prime}}\odot\mathbf{x}$, where ${{\bf{m}}_{\Omega^{\prime}}}= {\left[ {{m_1},{m_2}, \cdots,{m_M}} \right]^{\top}}$ is a masking vector with $m_{i} = 1$, if $i\in\Omega^{\prime}$ or $m_{i} = 0$ if $i\notin\Omega^{\prime}$.

All $4M_t M_r$ antennas are equipped with low-resolution ADCs. Therefore, the received measurements undergo uniform scalar quantization (scalar quantization with uniform dithering, i.e., $\boldsymbol{\uptau}=\left[\tau_i\sim \mathcal{U}_{\left[-\frac{\Delta}{2},\frac{\Delta}{2}\right]}+\mathrm{j}\mathcal{U}_{\left[-\frac{\Delta}{2},\frac{\Delta}{2}\right]}\right]$) as follows:
\begin{equation}
\label{h}
\widehat{x}_i=\mathcal{Q}_{_{\Delta}}\left(\operatorname{Re}\left(\bar{x}_i\right)\right)+\mathrm{j}\mathcal{Q}_{_{\Delta}}\left(\operatorname{Im}\left(\bar{x}_i\right)\right),\quad i\in\Omega^{\prime}.
\end{equation}
Considering that the quantization operator is applied solely to entries following $i\in\Omega^{\prime}$, we can express \eqref{h} in vector form as:
\begin{equation}
\label{f}
\widehat{\mbx} = \mathcal{Q}_{_{\Delta}}\left(\operatorname{Re}\left(\bar{\mbx}\right)\right)+\mathrm{j}\mathcal{Q}_{_{\Delta}}\left(\operatorname{Im}\left(\bar{\mbx}\right)\right),
\end{equation}
where $\widehat{\mbx}\in\mathcal{A}^{M}_{K}+\mathrm{j}\mathcal{A}^{M}_{K}$. Interestingly in \eqref{f}, the uniform quantizer becomes a one-bit quantizer by restricting the measurement's dynamic range to the scale parameter of uniform dithers as follows \cite{10414385}:
\begin{equation}
\begin{aligned}
&\widehat{\mbx}=\frac{\Delta}{2}\left[\operatorname{sgn}\left(\operatorname{Re}\left(\bar{\mbx}+\boldsymbol{\uptau}\right)\right)+\mathrm{j}\operatorname{sgn}\left(\operatorname{Im}\left(\bar{\mbx}+\boldsymbol{\uptau}\right)\right)\right],\\&\sup_{i\in [M_r M_t]}\left|\operatorname{Re}\left(\bar{x}_i\right)\right|,\sup_{i\in [M_r M_t]}\left|\operatorname{Im}\left(\bar{x}_i\right)\right|\leq\frac{\Delta}{2},
\end{aligned}
\end{equation}
where $\widehat{\mbx}\in\left\{-\frac{\Delta}{2},0,\frac{\Delta}{2}\right\}^{M}+\mathrm{j}\left\{-\frac{\Delta}{2},0,\frac{\Delta}{2}\right\}^{M}$.
In this paper, we partition the measurements with $i\in\Omega^{\prime}$, into two disjoint sets, $\Omega^{\prime}_1$ with $\left|\Omega^{\prime}_1\right|=M_{o}$
and $\Omega^{\prime}_2$ with $\left|\Omega^{\prime}_2\right|=M_{q}$ such that $M_{o}+M_{q}= 4M_tM_r$. Each measurement within the set $\Omega^{\prime}_1$ undergoes quantization into one-bit data using uniform dithering with a scale parameter $\Delta_1$, where $\Delta_1$ dominates the dynamic range of measurements associated with $i\in\Omega^{\prime}_1$. Each measurement within the set $\Omega^{\prime}_2$ undergoes quantization using uniform dithering with a scale parameter $\Delta_2$, where $\Delta_2$ is unable to dominate the dynamic range of the measurements associated with $i\in\Omega^{\prime}_2$. Hence, we have established a mixed-precision scenario in which $M_o$ entries are one-bit, while the rest are multi-bit. More generally, we define 
an indicator vector $\boldsymbol{\delta}=[\delta_1,\cdots, \delta_M]\in\{0,1\}^{M}$,
where for $i\in\Omega^{\prime}$, $\delta_i=1$ signifies that the $i$-th antenna is equipped with a high-precision ADC.
Therefore, the mixed-precision output can be represented as
\begin{equation}
\label{rep}
\mby = \operatorname{diag}\left(\boldsymbol{\delta}\right)\widehat{\mbx}+\left(\mbI-\operatorname{diag}\left(\boldsymbol{\delta}\right)\right)\frac{\Delta_1}{2}\mbr,
\end{equation}
where 
\begin{equation}
\begin{aligned}
\widehat{\mbx}&=\mathcal{Q}_{_{\Delta_2}}\left(\operatorname{Re}\left(\bar{\mbx}\right)\right)+\mathrm{j}\mathcal{Q}_{_{\Delta_2}}\left(\operatorname{Im}\left(\bar{\mbx}\right)\right),\\\Delta_2&<2\min\left(\sup_{i\in\Omega^{\prime}_2}|\operatorname{Re}\left(\bar{x}_i\right)|,\sup_{i\in\Omega^{\prime}_2}|\operatorname{Im}\left(\bar{x}_i\right)|\right),
\end{aligned}
\end{equation}
and
\begin{equation}
\begin{aligned}
\mbr&=\operatorname{sgn}\left(\operatorname{Re}\left(\bar{\mbx}-\boldsymbol{\uptau}\right)\right)+\mathrm{j}\operatorname{sgn}\left(\operatorname{Im}\left(\bar{\mbx}-\boldsymbol{\uptau}\right)\right),\\\Delta_1&\geq2\max\left(\sup_{i\in\Omega^{\prime}_1}|\operatorname{Re}\left(\bar{x}_i\right)|, \sup_{i\in\Omega^{\prime}_1}|\operatorname{Im}\left(\bar{x}_i\right)|\right).
\end{aligned}
\end{equation}
In numerous applications, the upper bound of the dynamic range of measurements is known \cite{cai2013max}. Therefore, based on this information, we can easily design the scale parameter of uniform dithers.

Following our model in \eqref{rep}, a Hankel matrix denoted by ${\mathcal{H}}\left( \mby \right) \in {\mathbb C}^{n_1 \times n_2}$ can be constructed as $[{\mathcal{H}}\left( \mby \right)]_{i,j} = y_{i+j-1}$, where ${n_{1} + n_{2}} = {M} + 1$. Specifically, in this paper, we adopt $n_1 = n_2 = \left({\frac{{M + 1}}{2}} \right)$ if $M$ is odd, and $n_1 = n_2 - 1 = \left( {\frac{{M}}{2}} \right) $ if $M$ is even. In \cite{sun20214d}, it has been demonstrated that the Hankel matrix possesses a low-rank property, with its rank equivalent to the number of targets. For the sake of simplicity in notation, hereafter, we denote the Hankel matrix of virtual sparse array response as ${\mathcal{H}}\left(\mbx\right)=\mbX$ and quantized data as ${\mathcal{H}}\left(\mby\right)=\mbQ$. 
In the rest of the paper, we will discuss the process of recovering $\mbX$ from $\mbQ$.

\subsection{Mixed-Quantized Hankel Matrix Completion}
\label{sec_3}
Assume $\Omega$ denotes the set of observed entries in $\mbX$ denoted by $\mathcal{P}_{\Omega}\left(\mbX\right)$, which is obtained by the array element indices of SLA.
In this section, we aim to obtain an upper bound for the recovery error in the mixed quantized complex matrix completion problem with high probability. Define the set of real low-rank matrices as
\begin{equation}
\label{real_lowrank}
\bar{\mathcal{K}}_r=\left\{\mbX^{\prime}\in\mathbb{R}^{n_1\times n_2}\mid\operatorname{rank}(\mbX^{\prime})\leq r, \|\mbX^{\prime}\|_{\mathrm{max}}\leq\alpha\right\}\subset\mathbb{R}^{n_1\times n_2}.
\end{equation}
Initially, we assume that $\mbX\in\bar{\mathcal{K}}_r$, and later, we extend the result to the complex scenario.
We will use the following definition in our theorem:
\begin{definition}
\label{def_1}
Define a low-rank matrix as $\mbX=[X_{i,j}]\in\bar{\mathcal{K}}_r$. The consistency property of uniform quantization over the pair $(\mbX,\mbY)\in\bar{\mathcal{K}}_r$, is given by
\begin{equation}
\label{a_1}
\mathcal{Q}_{_\Delta}\left(\mathcal{P}_{\Omega}\left(\mbX\right)\right)=\mathcal{Q}_{_\Delta}\left(\mathcal{P}_{\Omega}\left(\mbY\right)\right).
\end{equation}
\end{definition}
The concept of consistent reconstruction, as defined in Definition~\ref{def_1}, has played a central role in providing theoretical guarantees in the field of one-bit sensing \cite{jacques2017small, eamaz2023harnessing}. 
Before presenting our main recovery result, we establish an embedding between the metric spaces $\left(\bar{\mathcal{K}}_r\subset\mathbb{R}^{n_1\times n_2},\ell_1\right)$ and $\left(\mathcal{Q}_{_{\Delta}}\left(\mathcal{P}_{\Omega}\left(\bar{\mathcal{K}}_r\right)\right)\subset \mathcal{A}^{n_1\times n_2}_{K},\ell_1\right)$, where $\mathcal{Q}_{_{\Delta}}\left(\mathcal{P}_{\Omega}\left(\bar{\mathcal{K}}_r\right)\right)$ represents the space of quantized matrix completion. Specifically, the subsequent lemma establishes that the quantized mapping $\mathcal{Q}_{_{\Delta}}\left(\mathcal{P}_{\Omega}\left(\mbX\right)\right)$ is a quasi-isometric embedding between $\left(\bar{\mathcal{K}}_r,\ell_1\right)$ and $\left(\mathcal{Q}_{_{\Delta}}\left(\mathcal{P}_{\Omega}\left(\bar{\mathcal{K}}_r\right)\right),\ell_1\right)$:
\begin{lemma}
\label{lem_1}
For a pair $\left(\mbX,\mbY\right)\in\bar{\mathcal{K}}_r$, define the following distance:
\begin{equation}
D_{_{\Delta}}\left(\mbX,\mbY\right)=\frac{1}{m^{\prime}}\left\|\mathcal{Q}_{_\Delta}\left(\mathcal{P}_{\Omega}\left(\mbX\right)\right)-\mathcal{Q}_{_\Delta}\left(\mathcal{P}_{\Omega}\left(\mbY\right)\right)\right\|_1.
\end{equation}
Then, with a positive constant $\rho$, for all pairs $\left(\mbX,\mbY\right)\in\bar{\mathcal{K}}_r$, we have the following embedding:
\begin{equation}
\operatorname{Pr}\left(\sup_{\mbX,\mbY\in\bar{\mathcal{K}}_r}\left|D_{_{\Delta}}\left(\mbX,\mbY\right)-\frac{1}{n_1n_2}\left\|\mbX-\mbY\right\|_1\right|\leq\varepsilon\right)\leq 2 e^{-\frac{\epsilon^2 m^{\prime}}{K^2\Delta^2}},
\end{equation} 
as long as $m^{\prime}\gtrsim \varepsilon^{-2} r\left(n_1+n_2\right)\log\left(1+\frac{\|\bar{\mathcal{K}}_r\|}{\rho}\right)$.
\end{lemma}
In the mixed-quantized complex Hankel matrix completion approach, the observation of the partial matrix $\mathcal{P}_{\Omega}(\mbX)$ is made through $m^{\prime} = m^{\prime}_1 + m^{\prime}_2$ low-resolution samples, where $m^{\prime}_1$ are one-bit samples and $m^{\prime}_2$ are multi-bit samples. Here, $m^{\prime}$ is significantly smaller than $n_1n_2$. 
The mixed-precision Hankel matrix, denoted by $\mathbf{Q}$, is composed of entries at indices $(i,j) \in \Omega_1$—which are derived from $\Omega^{\prime}_1$ of the SLA data—containing one-bit data, and entries at indices $(i,j) \in \Omega_2$—also constructed based on $\Omega^{\prime}_2$ of SLA data—containing multi-bit data. Note that the union of $\Omega_1$ and $\Omega_2$ forms the complete set $\Omega$ of observed indices. The mixed-quantized matrix can be expressed as follows:
\begin{equation}
\label{Steph0}
\mbQ = \mathcal{Q}_{_{\Delta_1}}\left(\operatorname{Re}\left(\mathcal{P}_{\Omega_1}\left(\mbX\right)\right)\right)+\mathcal{Q}_{_{\Delta_2}}\left(\operatorname{Re}\left(\mathcal{P}_{\Omega_2}\left(\mbX\right)\right)\right)+\mathrm{j}\left(\mathcal{Q}_{_{\Delta_1}}\left(\operatorname{Im}\left(\mathcal{P}_{\Omega_1}\left(\mbX\right)\right)\right)+\mathcal{Q}_{_{\Delta_2}}\left(\operatorname{Im}\left(\mathcal{P}_{\Omega_2}\left(\mbX\right)\right)\right)\right).
\end{equation}

In this paper, we focus on the complex matrix completion problem. 
For the real part, it is straightforward to verify that the distance defined in Lemma~\ref{lem_1}, in the mixed quantization scheme, is given by
\begin{equation}
\begin{aligned}
\label{dis_1}
D_{_{\Delta_1,\Delta_2}}^{r}\left(\mbX,\mbY\right)&=\frac{\Delta_1}{2 m^{\prime}_1}\left\|\operatorname{sgn}\left(\operatorname{Re}\left(\mathcal{P}_{\Omega_1}\left(\mbX\right)\right)\right)-\operatorname{sgn}\left(\operatorname{Re}\left(\mathcal{P}_{\Omega_1}\left(\mbY\right)\right)\right)\right\|_1 \\&+\frac{1}{m^{\prime}_2}\left\|\mathcal{Q}_{_{\Delta_2}}\left(\operatorname{Re}\left(\mathcal{P}_{\Omega_2}\left(\mbX\right)\right)\right)-\mathcal{Q}_{_{\Delta_2}}\left(\operatorname{Re}\left(\mathcal{P}_{\Omega_2}\left(\mbY\right)\right)\right)\right\|_1.
\end{aligned}
\end{equation}
Similarly, for the imaginary part, we have
\begin{equation}
\begin{aligned}
\label{dis_2}
D_{_{\Delta_1,\Delta_2}}^{i}\left(\mbX,\mbY\right)&=\frac{\Delta_1}{2 m^{\prime}_1}\left\|\operatorname{sgn}\left(\operatorname{Im}\left(\mathcal{P}_{\Omega_1}\left(\mbX\right)\right)\right)-\operatorname{sgn}\left(\operatorname{Im}\left(\mathcal{P}_{\Omega_1}\left(\mbY\right)\right)\right)\right\|_1 \\&+\frac{1}{m^{\prime}_2}\left\|\mathcal{Q}_{_{\Delta_2}}\left(\operatorname{Im}\left(\mathcal{P}_{\Omega_2}\left(\mbX\right)\right)\right)-\mathcal{Q}_{_{\Delta_2}}\left(\operatorname{Im}\left(\mathcal{P}_{\Omega_2}\left(\mbY\right)\right)\right)\right\|_1.
\end{aligned}
\end{equation}
Following Lemma~\ref{lem_1}, we determine an embedding for each real and imaginary part with associated probabilities. Additionally, the triangle inequality implies that the error of a complex value can be bounded by the sum of its real and imaginary parts.
Furthermore, it is worth mentioning that the statistical properties of both real and imaginary values are identical. Consequently, employing the union bound, the resulting failure probability is doubled compared to the real part alone. 
The subsequent theorem provides an upper-bound for the recovery performance in mixed-precision matrix completion:
\begin{theorem}
\label{Theorem_q}
Let $\mathcal{K}_r$ denote the set of low-rank matrices defined as
\begin{equation}
\label{n_02}
\mathcal{K}_r=\left\{\mbX^{\prime}\in\mathbb{C}^{n_1\times n_2}\mid\operatorname{rank}(\mbX^{\prime})\leq r, \beta\leq\alpha\right\}\subset\mathbb{C}^{n_1\times n_2},
\end{equation}
where $\beta= \max\left(\|\operatorname{Re}\left(\mbX^{\prime}\right)\|_{\mathrm{max}},\|\operatorname{Im}\left(\mbX^{\prime}\right)\|_{\mathrm{max}}\right)$.
Consider a matrix $\mbX\in\mathcal{K}_r$. Assume that $m^{\prime}=m^{\prime}_1+m^{\prime}_2$ entries of $\mbX$, randomly selected with uniform sampling, undergo mixed-precision scalar quantization followed by \eqref{rep} with dither values following $\mathcal{U}_{[-\frac{\Delta_1}{2},\frac{\Delta_1}{2}]}+\mathrm{j}\mathcal{U}_{[-\frac{\Delta_1}{2},\frac{\Delta_1}{2}]}$ for $m^{\prime}_1$ one-bit data, and $\mathcal{U}_{[-\frac{\Delta_2}{2},\frac{\Delta_2}{2}]}+\mathrm{j}\mathcal{U}_{[-\frac{\Delta_2}{2},\frac{\Delta_2}{2}]}$ for $m^{\prime}_2$ multi-bit data. If the pair $\left(\mbX, \mbY\right)\in\mathcal{K}_r$ satisfies the consistency property in Definition~\ref{def_1} for each real and imaginary part, with constants $c$, $\left(\varepsilon_1,\varepsilon_2\right) \in (0,1)$, it can be asserted that the recovery error between $\mbX$ and $\mbY$ is bounded with a probability of at least $1-4\max\left(e^{-\frac{\varepsilon^2_1 m^{\prime}_1}{\Delta^2_1}},e^{-\frac{\varepsilon^2_2 m^{\prime}_2}{K^2\Delta^2_2}}\right)$ as
\begin{equation}
\label{a_50}
\left\|\mbX-\mbY\right\|_{1}\leq 2 n_1 n_2 \left(\varepsilon_1+\varepsilon_2\right),
\end{equation}
for all $\left(\mbX, \mbY\right) \in\mathcal{K}_r$ when the required number of samples must satisfy
\begin{equation}
m^{\prime}\gtrsim \min\left(\varepsilon_1, \varepsilon_2\right)^{-2} r\left(n_1+n_2\right)\log\left(1+\frac{\|\mathcal{K}_r\|}{\rho}\right),
\end{equation}
with $\rho\leq2 n_1 n_2 \left(\varepsilon_1+\varepsilon_2\right)$.
\end{theorem}
The proof of Theorem~\ref{Theorem_q} is presented in Appendix~\ref{proof0}.

To recover the matrix $\mbX$ from the observed mixed-quantized matrix $\mbQ$, we employ the nuclear norm minimization problem:
\begin{equation}
\label{Steph3}
\begin{aligned}
&\underset{\mbX}{\textrm{minimize}}\quad \left\|\mbX\right\|_{\star}\\
&\text{subject to} \quad \left\|\mathcal{P}_{\Omega}(\mbX)-\mbQ\right\|_{\mathrm{F}} \leq q,
\end{aligned}
\end{equation}
where the parameter $q$ denotes the impact of the quantization process. 

Denote a linear transformation by $\mathcal{M}: \mathbb{C}^{n_1\times n_2}\rightarrow \mathbb{C}^{m^{\prime}}$ and its adjoint operator as $\mathcal{M}^{\star}:\mathbb{C}^{m^{\prime}}\rightarrow \mathbb{C}^{n_1\times n_2}$. We employ the singular value thresholding (SVT) algorithm to address the nuclear norm minimization problem in mixed-precision Hankel matrix completion. If we consider the singular value decomposition (SVD) of $\mbX$ as $\mbX=\mbU\bSigma\mbV^{\top}$ and $\{\sigma_i\}$ as its singular values, the SVT uses the singular value shrinkage operator  which applies the partial SVD
to achieve the low-rank matrix structure as 
$\mathcal{D}_\tau(\mbX)=\mbU \mathcal{D}_\tau(\bSigma) \mbV^{\top},~ \mathcal{D}_\tau(\boldsymbol{\Sigma})=\operatorname{diag}\left(\left(\sigma_i-\tau\right)^{+}\right)$,
where $\tau\geq 0$ is the predefined threshold. The key distinction lies in our approach: rather than utilizing high-resolution partial measurements at each iteration, we employ mixed-precision data in the following manner:
\begin{equation}
\label{St_22}
\left\{\begin{array}{l}
\mbX^{(k)}=\mathcal{D}_\tau\left(\mathcal{M}^{\star}\left(\boldsymbol{y}^{(k-1)}\right)\right),\\
\boldsymbol{y}^{(k)}=\boldsymbol{y}^{(k-1)}+\delta_k \left(\mbb-\mathcal{M}\left(\mbX^{(k)}\right)\right),
\end{array}\right.
\end{equation} 
where $\left\{\delta_k\right\}$ are step sizes and $\mbb=\operatorname{vec}\left([Q_{i,j}], ~{(i,j)\in \Omega}\right)$. 

\section{Numerical Investigation}
In this section, we numerically scrutinize the performance of the one-dimensional sparse array completion that jointly utilizes sparse spectrum and sparse arrays with mixed-precision measurements for radar collaborative sensing in automotive applications.


To attain high azimuth angular resolution, we cascade multiple automotive radar transceivers to synthesize a large sparse array in azimuth. We focus on the same physical array illustrated in Fig.~\ref{collab_sensing} (c), where there are $M_t=3$ transmit antennas, and $M_r=4$ receive antennas for each radar arranged along the horizontal direction at:
\begin{equation}
\begin{aligned}
& l_{\mathrm{TX_{1}}}=[1,9,25] \lambda / 2, \\
& l_{\mathrm{TX_{2}}}=[51,67,75] \lambda / 2, \\
& l_{\mathrm{RX_{1}}}=[1,6,7,8] \lambda / 2, \\
& l_{\mathrm{RX_{2}}}=[68,69,70,75] \lambda / 2.
\end{aligned}
\end{equation}
Here, $l_{\mathrm{TX_{1}}}=[1,9,25]$ and $l_{\mathrm{RX_{1}}}=[1,6,7,8]$ denote the Tx and Rx array element positions for radar 1, while $l_{\mathrm{TX_{2}}}=[51,67,75]$ and $l_{\mathrm{RX_{2}}}=[68,69,70,75]$ represent the Tx and Rx array element positions for radar 2. In this configuration, the first Tx and Rx antennas of the radar share the same position, resulting in the virtual array size of $4MtMr$. In addition, two virtual array elements overlap, which makes the final array element number $4MtMr-1$. Therefore, a virtual array with a total of $47$ elements is synthesized. The transmit and receive antennas and the virtual array are plotted
in Fig.~\ref{collab_sensing} (c). The azimuth angles are $\theta_1 = -34^{\circ}$ and $\theta_2 = 18^{\circ}$. Our simulations set the SNR as $20$~$\mathrm{dB}$. Initially, the two targets can be first separated in range-Doppler. The complex peak values in the range-Doppler spectrum, corresponding to each virtual sparse array, constitute an array snapshot for azimuth angle determination.

Following Section~\ref{sec_1}, we have considered the mixed-precision scenario with $44$ one-bit ADCs and varying placements of four 10-bit ADCs. Specifically, three different configurations have been considered for the $10$-bit ADCs: (a) four multi-bit ADCs are uniformly positioned along the edge of linear arrays, (b) the four multi-bit ADCs are allocated to the last four antennas in the sequence, and (c) the initial four antennas in the sequence are equipped with the multi-bit ADCs. For each configuration,
a Hankel matrix $\mbQ\in\mathbb{C}^{75\times 75}$ is constructed based on the array response of a ULA with $150$ elements and sampled by the SLA shown in Fig.~\ref{collab_sensing} (c). In the constructed Hankel matrix, $33$ percent of elements are non-zero while $1871$ non-zero elements are one-bit and $22$ of non-zero elements are $10$-bit values. Hence, the mixed-precision rate, defined as the ratio of multi-bit data to one-bit data, is $0.0118$.
\begin{figure*}[t]
	\centering
	\subfigure[]
		{\includegraphics[width=0.3\columnwidth]{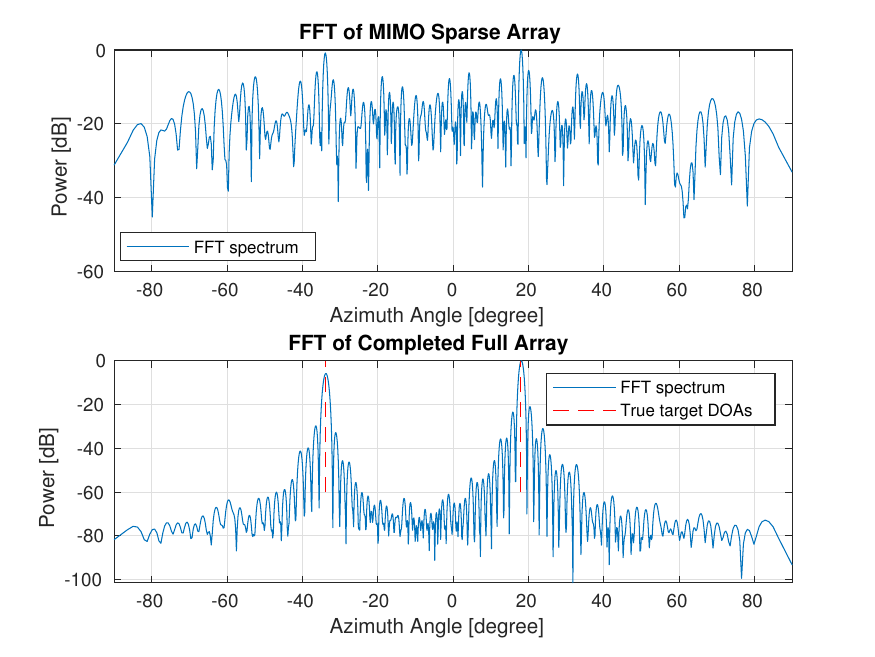}}\qquad
    \subfigure[]
		{\includegraphics[width=0.3\columnwidth]{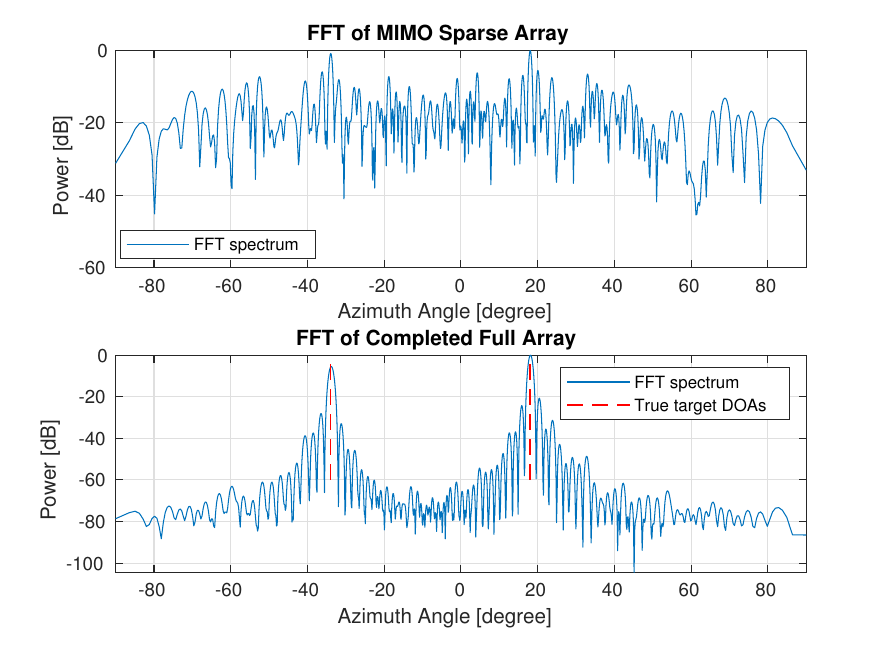}}\qquad
    \subfigure[]
		{\includegraphics[width=0.3\columnwidth]{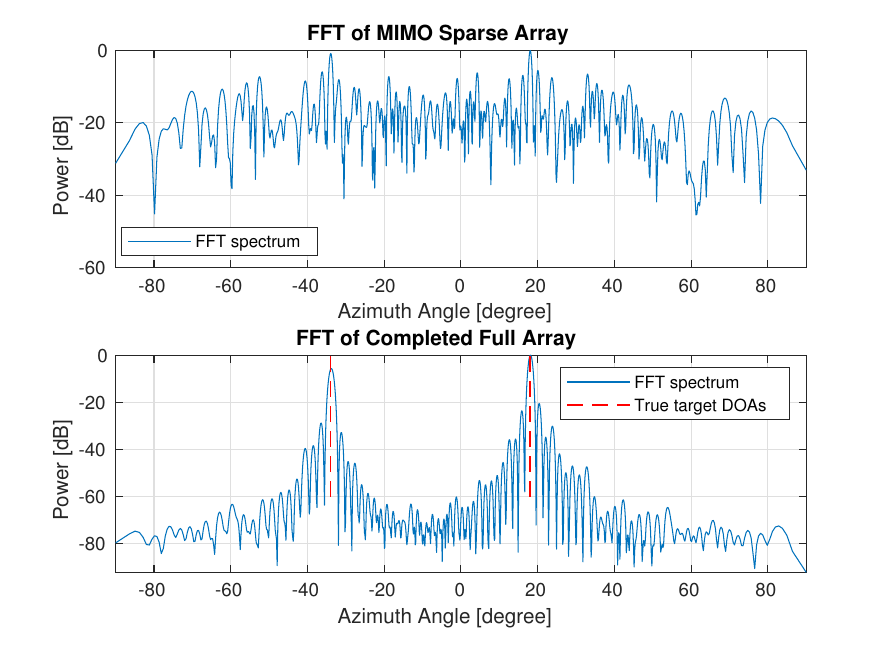}}\qquad
	\caption{Comparison between the spectrum of two targets with azimuth angles of $\{\theta_1,\theta_2\} = \{-34^{\circ}, 18^{\circ}\}$ under MIMO sparse array and fully completed array from mixed-quantized Hankel matrix employing 44 one-bit ADCs, along with varying placements of four 10-bit ADCs, is explored in three distinct configurations:
(a) Four multi-bit ADCs are uniformly positioned along the edge of linear arrays, with two located at the beginning and two at the end of the antenna sequence.
(b) The four multi-bit ADCs are allocated to the last four antennas in the sequence.
(c) The initial four antennas in the sequence are equipped with the multi-bit ADCs.
}
\vspace{-13pt}
\label{figure_1}
\end{figure*}

Following Section~\ref{sec_3}, the rank-$2$ Hankel matrix $\mbX$ is completed via the proposed algorithm in \eqref{St_22}. Let $\bar{\mbX}$ denote the completed Hankel matrix. Then, the full ULA response can be reconstructed by taking the average of the anti-diagonal elements of the matrix $\bar{\mbX}$. The completed full array has an aperture size of $76\lambda$.

Fig.~\ref{figure_1} depicts
the angle spectrum for the two targets in each configuration. The azimuth angle spectra are derived by applying FFT to the original SLA with the holes filled with zeros and the full array completed via mixed-quantized matrix completion, respectively. The FFT of the SLA produces two peaks corresponding to the correct azimuth directions but with high sidelobes, making it challenging to detect the two targets accurately in azimuth. In contrast, the completed full array from mixed-precision samples exhibits two distinct peaks corresponding to the correct azimuth locations in the angle spectrum, with significantly suppressed sidelobes. As can be seen in Fig.~\ref{figure_1}, the ADCs arrangement does not contribute significantly to the accuracy of the recovered azimuth angles following the mixed-quantized Hankel matrix completion approach.
\begin{figure*}[t]
	\centering
	\subfigure[]
		{\includegraphics[width=0.3\columnwidth]{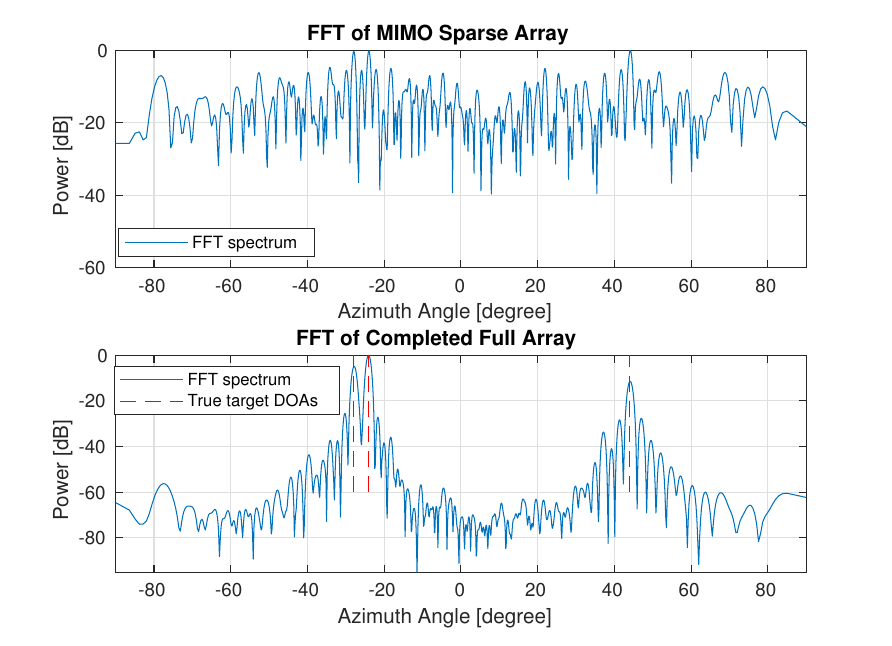}}\qquad
    \subfigure[]
		{\includegraphics[width=0.3\columnwidth]{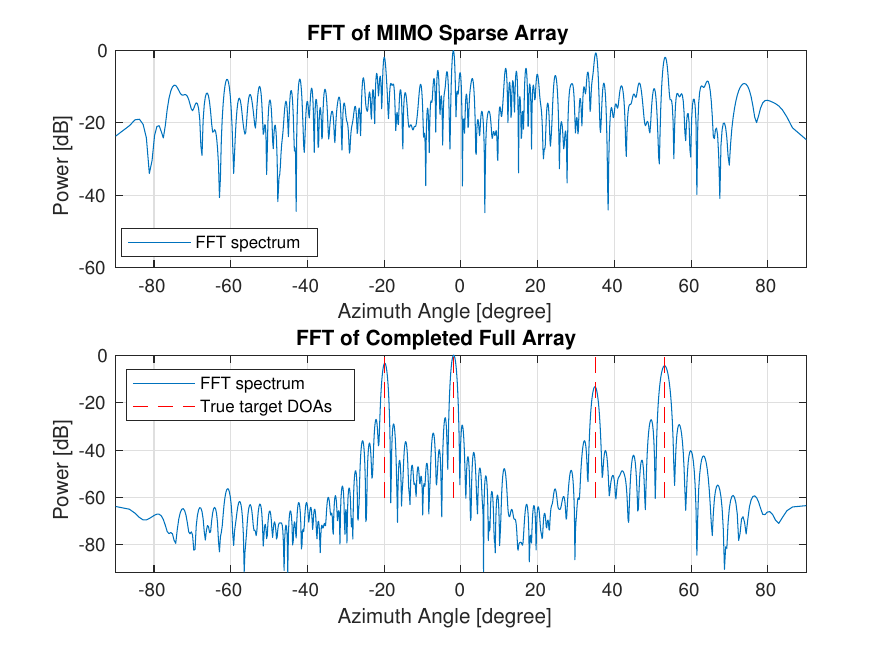}}\qquad
    \subfigure[]
		{\includegraphics[width=0.3\columnwidth]{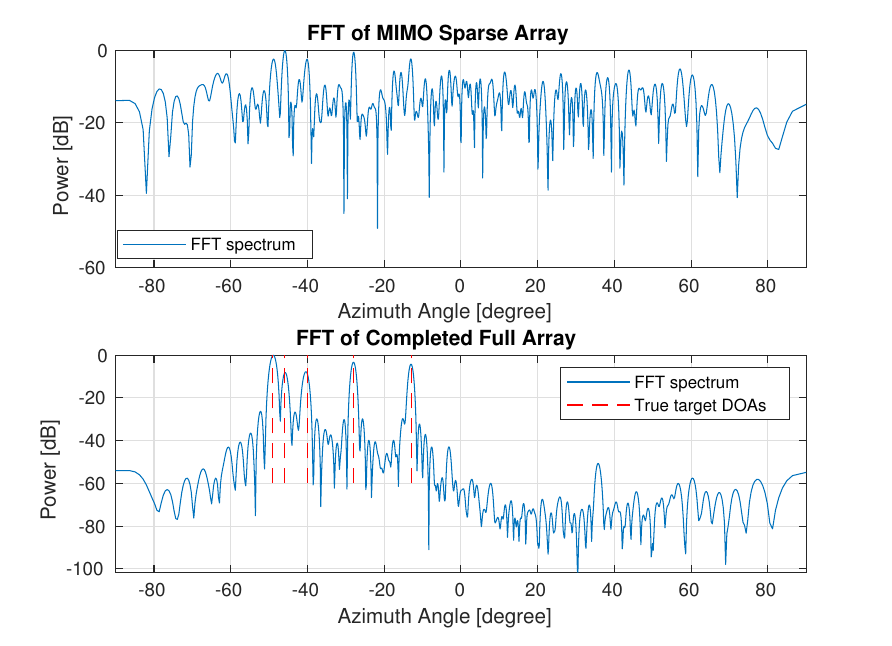}}\qquad
	\caption{The spectrum of (a) three targets with azimuth angles of $\{\theta_1,\theta_2,\theta_3\} = \{-28^{\circ}, -24^{\circ},44^{\circ}\}$, (b) four targets with azimuth angles of $\{\theta_1,\theta_2,\theta_3,\theta_4\} = \{-20^{\circ}, -2^{\circ},35^{\circ},53^{\circ}\}$, and five targets with azimuth angles of $\{\theta_1,\theta_2,\theta_3,\theta_4,\theta_5\}=\{-49^{\circ}, -46^{\circ},-40^{\circ},-28^{\circ},-13^{\circ}\}$, under MIMO sparse array and fully completed array from mixed-precision data.}
\vspace{-15pt}
\label{figure_2}
\end{figure*}

In our next experiments, we explore scenarios involving three, four, and five targets, each exhibiting different azimuth angles, using the same configuration as considered in Fig.~\ref{figure_1} (c).
It is important to emphasize that we consider the same number of one-bit and $10$-bit entries based on this configuration while trying to recover the rank-$3$,$4$, and $5$ Hankel matrices $\mbX$. As can be observed in Figs.~\ref{figure_2} (a)-(c), we can successfully recover the azimuth angles from the mixed-quantized data.
\section{Discussion}
We integrated mixed-precision ADCs into an SLA synthesized by a collaborative radar sensing paradigm employed in high-resolution automotive imaging radar applications. By implementing a properly designed uniform dithering scheme, we achieved precise target detection in azimuth while notably suppressing spectrum sidelobes. Our empirical findings indicated that the arrangement of multi-bit and one-bit ADCs has minimal impact on Hankel matrix completion and azimuth detection performance, yielding nearly identical results. However, a remaining question for future exploration is the theoretical investigation of this observed indifference in Hankel matrix completion performance concerning the mixed ADC arrangement.
\appendices
\section{Proof of Theorem~\ref{Theorem_q}}
\label{proof0}
Consider the set of real low-rank matrices $\bar{\mathcal{K}}_r$ as defined in \eqref{real_lowrank}. Initially, we focus on quantized matrix completion given $\mathcal{Q}_{_{\Delta}}\left(\mathcal{P}_{\Omega}\left(\mbX\right)\right)$ for any $\mbX\in\bar{\mathcal{K}}_r$. Subsequently, with slight adjustments, we extend the proof to mixed quantized complex matrix completion. 
Consider the following lemma for the scalar quantizer:
\begin{lemma}{\cite[Appendix~A]{jacques2017small}}
\label{Jack}
For real values $a, b\in\mathbb{R}$ and $\tau\sim\mathcal{U}_{[-\frac{\Delta}{2},\frac{\Delta}{2}]}$, we have
\begin{equation}
\mathbb{E}_{\tau}\left\{\left|\mathcal{Q}_{_{\Delta}}(a+\tau)-\mathcal{Q}_{_{\Delta}}(b+\tau)\right|\right\} = \left|a-b\right|.
\end{equation}
\end{lemma}
According to the aforementioned lemma, for any pair $\left(\mbX,\mbY\right)\in\bar{\mathcal{K}}_r$, the following $\ell_1$-distance equation between the qunatized values holds true:
\begin{lemma}
\label{yuta}
If we have the uniform sampling in the matrix completion problem, for a pair $\left(\mbX,\mbY\right)\in\bar{\mathcal{K}}_r$ and dither values follow $\mathcal{U}_{[-\frac{\Delta}{2},\frac{\Delta}{2}]}$, the following relation is obtained:
\begin{equation}
\mathbb{E}_{\tau,(i,j)}\left\{\left\|\mathcal{Q}_{_{\Delta}}\left(\mathcal{P}_{\Omega}\left(\mbX\right)\right)-\mathcal{Q}_{_{\Delta}}\left(\mathcal{P}_{\Omega}\left(\mbY\right)\right)\right\|_1\right\} = \frac{m^{\prime}}{n_1n_2}\left\|\mbX-\mbY\right\|_1.
\end{equation}
\end{lemma}
\begin{IEEEproof}
The proof of Lemma~\ref{yuta}, is straightforward by considering Lemma~\ref{Jack} for the expectation over the dither values. Since we use the uniform sampling in the matrix completion, the expectation over the indices $(i,j)\in\Omega$ is readily given by
\begin{equation}
\mathbb{E}_{(i,j)}\left\{\sum_{(i,j)\in\Omega}\left|X_{i,j}-Y_{i,j}\right|\right\} = \sum_{(i,j)\in\Omega}\sum_{(i,j)\in[n_1]\times[n_2]}\frac{1}{n_1n_2}\left|X_{i,j}-Y_{i,j}\right| = \frac{m^{\prime}}{n_1n_2}\left\|\mbX-\mbY\right\|_1.  
\end{equation}
\end{IEEEproof}
Herein, we want to obtain the probability of how the $\ell_1$-distance of the quantized values concentrates around its mean by using Hoeffding's inequality for the bounded random variable:
\begin{lemma}\cite[Theroem~2.2.5]{vershynin2018high}
\label{lemma_hoeffding} 
Let $\left\{X_i\right\}^{n}_{i=1}$ be independent, bounded random variables
satisfying $X_i \in [a_i, c_i]$, then for any $t > 0$ it holds that 
\begin{equation}
\operatorname{Pr}\left(\left|\frac{1}{n} \sum_{i=1}^n\left(X_i-\mathbb{E}\left\{ X_i\right\}\right)\right| \geq t\right)\leq 2 e^{-\frac{2 n^2 t^2}{\sum_{i=1}^n\left(c_i-a_i\right)^2}}.  
\end{equation}
\end{lemma}
Also, we have
\begin{equation}
\left|\mathcal{Q}_{_{\Delta}}(X_{i,j})-\mathcal{Q}_{_{\Delta}}(Y_{i,j})\right|\leq K\Delta,\quad (i,j)\in\Omega.
\end{equation}
Therefore, we can write the following concentration inequality:
\begin{equation}
\label{Kenjaku}
\operatorname{Pr}\left(\left|\frac{1}{m^{\prime}}\left\|\mathcal{Q}_{_{\Delta}}\left(\mathcal{P}_{\Omega}\left(\mbX\right)\right)-\mathcal{Q}_{_{\Delta}}\left(\mathcal{P}_{\Omega}\left(\mbY\right)\right)\right\|_1-\frac{1}{n_1n_2}\left\|\mbX-\mbY\right\|_1\right|\leq\varepsilon\right)\leq 2 e^{-\frac{2 \varepsilon^2 m^{\prime}}{K^2\Delta^2}}.  
\end{equation}
We take a finite covering of $\bar{\mathcal{K}}_r$ by a $\rho$-net $\mathcal{G}_{\rho}\subset \bar{\mathcal{K}}_r$ (for $0<\rho\leq\varepsilon$) and we extend the concentration of $\ell_1$-distance to all vectors of $\mathcal{G}_{\rho}$ by using the following fact:
\begin{equation}
\label{amu_arian_1}
\log\left(|\mathcal{G}_{\rho}|\right) = \mathcal{H}\left(\bar{\mathcal{K}}_r,\rho\right).   
\end{equation}
For structured sets, the Kolmogorov $\rho$-entropy is upper bounded as \cite{jacques2017small}
\begin{equation}
\label{amu_arian_2}
\mathcal{H}\left(\bar{\mathcal{K}}_r,\rho\right)\leq \gamma^2\left(\bar{\mathcal{K}}_r\right)\log\left(1+\frac{\|\bar{\mathcal{K}}_r\|}{\rho}\right), 
\end{equation}
where the gaussian complexity is upper-bounded for $\bar{\mathcal{K}}_r$ as
\begin{equation}
\label{amu_arian_3}
\gamma^{2}\left(\bar{\mathcal{K}}_r\right) \leq r\left(n_1+n_2\right).  
\end{equation}
Consequently, we have
\begin{equation}
\label{Sukuna}
\operatorname{Pr}\left(\sup_{\mbX,\mbY\in\bar{\mathcal{K}}_r}\left|\frac{1}{m^{\prime}}\left\|\mathcal{Q}_{_{\Delta}}\left(\mathcal{P}_{\Omega}\left(\mbX\right)\right)-\mathcal{Q}_{_{\Delta}}\left(\mathcal{P}_{\Omega}\left(\mbY\right)\right)\right\|_1-\frac{1}{n_1n_2}\left\|\mbX-\mbY\right\|_1\right|\leq\varepsilon\right)\leq 2 \left|\mathcal{G}_{\rho}\right|^2e^{-\frac{2 \varepsilon^2 m^{\prime}}{K^2\Delta^2}},  
\end{equation}
or equivalently,
\begin{equation}
\label{yuji}
\operatorname{Pr}\left(\sup_{\mbX,\mbY\in\bar{\mathcal{K}}_r}\left|\frac{1}{m^{\prime}}\left\|\mathcal{Q}_{_{\Delta}}\left(\mathcal{P}_{\Omega}\left(\mbX\right)\right)-\mathcal{Q}_{_{\Delta}}\left(\mathcal{P}_{\Omega}\left(\mbY\right)\right)\right\|_1-\frac{1}{n_1n_2}\left\|\mbX-\mbY\right\|_1\right|\leq\varepsilon\right)\leq 2 e^{2\mathcal{H}\left(\bar{\mathcal{K}}_r,\rho\right)-\frac{2 \varepsilon^2 m^{\prime}}{K^2\Delta^2}},
\end{equation}
which leads to obtain the required number of samples as follows:
\begin{equation}
\label{nobara}
m^{\prime}\gtrsim \varepsilon^{-2} r\left(n_1+n_2\right)\log\left(1+\frac{\|\bar{\mathcal{K}}_r\|}{\rho}\right).
\end{equation}
By considering the concentration inequality \eqref{yuji}, with probability exceeding $1-e^{-\frac{\varepsilon^2 m^{\prime}}{K^2\Delta^2}}$ we can write
\begin{equation}
\frac{1}{n_1n_2} \left\|\mbX-\mbY\right\|_1-\varepsilon\leq \frac{1}{m^{\prime}}\left\|\mathcal{Q}_{_{\Delta}}\left(\mathcal{P}_{\Omega}\left(\mbX\right)\right)-\mathcal{Q}_{_{\Delta}}\left(\mathcal{P}_{\Omega}\left(\mbY\right)\right)\right\|_1\leq \frac{1}{n_1n_2} \left\|\mbX-\mbY\right\|_1+\varepsilon,
\end{equation}
or equivalently,
\begin{equation}
\label{gojo}
\left\|\mbX-\mbY\right\|_1\leq \frac{n_1 n_2}{m^{\prime}}\left\|\mathcal{Q}_{_{\Delta}}\left(\mathcal{P}_{\Omega}\left(\mbX\right)\right)-\mathcal{Q}_{_{\Delta}}\left(\mathcal{P}_{\Omega}\left(\mbY\right)\right)\right\|_1+n_1n_2 \varepsilon.  
\end{equation}
Now, we want to extend the proof for the pair $(\mbX,\mbY)\in\mathcal{K}_r$ while considering the mixed quantization scheme. According to Lemma~\ref{yuta}, for the distance
$D_{_{\Delta_1,\Delta_2}}^r$ defined in \eqref{dis_1}, we can write
\begin{equation}
\label{yutaa}
\begin{aligned}
\mathbb{E}_{\tau,(i,j)}\left\{\left\|\mathcal{Q}_{_{\Delta_1}}\left(\operatorname{Re}\left(\mathcal{P}_{\Omega_1}\left(\mbX\right)\right)\right)-\mathcal{Q}_{_{\Delta_1}}\left(\operatorname{Re}\left(\mathcal{P}_{\Omega_1}\left(\mbY\right)\right)\right)\right\|_1\right\} &= \frac{m^{\prime}_1}{n_1n_2}\left\|\operatorname{Re}\left(\mbX-\mbY\right)\right\|_1,\\
\mathbb{E}_{\tau,(i,j)}\left\{\left\|\mathcal{Q}_{_{\Delta_2}}\left(\operatorname{Re}\left(\mathcal{P}_{\Omega_2}\left(\mbX\right)\right)\right)-\mathcal{Q}_{_{\Delta_2}}\left(\operatorname{Re}\left(\mathcal{P}_{\Omega_2}\left(\mbY\right)\right)\right)\right\|_1\right\} &= \frac{m^{\prime}_2}{n_1n_2}\left\|\operatorname{Re}\left(\mbX-\mbY\right)\right\|_1,
\end{aligned}
\end{equation}
with $\alpha=2\Delta_1$. Similarly for $D_{_{\Delta_1,\Delta_2}}^i$ defined in \eqref{dis_2}, we can write
\begin{equation}
\label{yuta_1}
\begin{aligned}
\mathbb{E}_{\tau,(i,j)}\left\{\left\|\mathcal{Q}_{_{\Delta_1}}\left(\operatorname{Im}\left(\mathcal{P}_{\Omega_1}\left(\mbX\right)\right)\right)-\mathcal{Q}_{_{\Delta_1}}\left(\operatorname{Im}\left(\mathcal{P}_{\Omega_1}\left(\mbY\right)\right)\right)\right\|_1\right\} &= \frac{m^{\prime}_1}{n_1n_2}\left\|\operatorname{Im}\left(\mbX-\mbY\right)\right\|_1,\\
\mathbb{E}_{\tau,(i,j)}\left\{\left\|\mathcal{Q}_{_{\Delta_2}}\left(\operatorname{Im}\left(\mathcal{P}_{\Omega_2}\left(\mbX\right)\right)\right)-\mathcal{Q}_{_{\Delta_2}}\left(\operatorname{Im}\left(\mathcal{P}_{\Omega_2}\left(\mbY\right)\right)\right)\right\|_1\right\} &= \frac{m^{\prime}_2}{n_1n_2}\left\|\operatorname{Im}\left(\mbX-\mbY\right)\right\|_1.
\end{aligned}
\end{equation}
Consequently, for $D_{_{\Delta_1,\Delta_2}}^r$,
the following concentration inequality is obtained:
\begin{equation}
\label{t1}
\operatorname{Pr}\left(\sup_{\mbX,\mbY\in\mathcal{K}_r}\left|D^r_{_{\Delta_1,\Delta_2}}\left(\mbX,\mbY\right)-\frac{1}{n_1n_2}\left\|\operatorname{Re}\left(\mbX-\mbY\right)\right\|_1\right|\leq(\varepsilon_1+\varepsilon_2)\right)\leq 2 \max\left(e^{-\frac{\varepsilon^2_1 m^{\prime}_1}{\Delta^2_1}},e^{-\frac{\varepsilon^2_2 m^{\prime}_2}{K^2\Delta^2_2}}\right),
\end{equation} 
where $\varepsilon_1$ is associated with the concentration inequality of one-bit data and $\varepsilon_2$ is related to the multi-bit data. Similarly for $D_{_{\Delta_1,\Delta_2}}^i$, we have 
\begin{equation}
\label{t2}
\operatorname{Pr}\left(\sup_{\mbX,\mbY\in\mathcal{K}_r}\left|D^i_{_{\Delta_1,\Delta_2}}\left(\mbX,\mbY\right)-\frac{1}{n_1n_2}\left\|\operatorname{Im}\left(\mbX-\mbY\right)\right\|_1\right|\leq(\varepsilon_1+\varepsilon_2)\right)\leq 2 \max\left(e^{-\frac{\varepsilon^2_1 m^{\prime}_1}{\Delta^2_1}},e^{-\frac{\varepsilon^2_2 m^{\prime}_2}{K^2\Delta^2_2}}\right).
\end{equation}
In the consistency scenario, we have
\begin{equation}
\begin{aligned}
\mathcal{Q}_{_{\Delta_{1}}}\left(\operatorname{Re}\left(\mathcal{P}_{\Omega_1}\left(\mbX\right)\right)\right)&=\mathcal{Q}_{_{\Delta_{1}}}\left(\operatorname{Re}\left(\mathcal{P}_{\Omega_1}\left(\mbY\right)\right)\right),\\\mathcal{Q}_{_{\Delta_{2}}}\left(\operatorname{Re}\left(\mathcal{P}_{\Omega_2}\left(\mbX\right)\right)\right)&=\mathcal{Q}_{_{\Delta_{2}}}\left(\operatorname{Re}\left(\mathcal{P}_{\Omega_2}\left(\mbY\right)\right)\right).
\end{aligned}
\end{equation}
A similar result is true for the imaginary part. Combining the consistency result with \eqref{t1} and \eqref{t2} leads to
\begin{equation}
\left\|\operatorname{Re}\left(\mbX-\mbY\right)\right\|_1+\left\|\operatorname{Im}\left(\mbX-\mbY\right)\right\|_1\leq2n_1n_2(\varepsilon_1+\varepsilon_2),
\end{equation}
where the triangle inequality implies
\begin{equation}
\|\mbX-\mbY\|_{\mathrm{F}}\leq 2n_1n_2(\varepsilon_1+\varepsilon_2).
\end{equation}
Note that the factor of $4$ in the probability arises from the fact that both real and imaginary parts have the same probability. According to the union bound, the overall probability is twice as large. Setting $\rho\leq2n_1n_2(\varepsilon_1+\varepsilon_2)$ completes the proof.
\bibliographystyle{IEEEtran}
\bibliography{references}

\end{document}

%% file: symbols.tex
\usepackage[dvipsnames]{xcolor}
\usepackage{amsmath}
\usepackage{amsthm}
\usepackage{dsfont}
\usepackage{thmtools}
\usepackage{amssymb}

\def\bSigma{\boldsymbol{\Sigma}}

\def\mbb{\mathbf{b}}

\def\mbr{\mathbf{r}}

\def\mbx{\mathbf{x}}
\def\mby{\mathbf{y}}

\def\mbI{\mathbf{I}}

\def\mbQ{\mathbf{Q}}

\def\mbU{\mathbf{U}}
\def\mbV{\mathbf{V}}

\def\mbX{\mathbf{X}}
\def\mbY{\mathbf{Y}}

\newtheorem{theorem}{Theorem}

\newtheorem{lemma}{Lemma}

\theoremstyle{definition}
\newtheorem{definition}{Definition}

